\documentstyle[aps,psfig]{revtex}
\begin{document}
\title{New Dark Matter Physics: Clues from Halo Structure}
\author{Craig J. Hogan and Julianne J. Dalcanton}
\address{Astronomy  Department, 
University of Washington,
Seattle, Washington 98195-1580}

\maketitle

\begin{abstract}
We examine the effect of  primordial dark matter
velocity dispersion and/or particle self-interactions on the structure and stability
of galaxy halos, especially with respect to the formation of  substructure
and central density cusps. 
Primordial velocity dispersion  is characterised by a ``phase density''
$Q\equiv \rho/\langle v^2\rangle^{3/2}$,  which
 for relativistically-decoupled relics is determined by   particle mass and
spin and is  insensitive to cosmological parameters.   
Finite $Q$
 leads  to small-scale  filtering    of  the primordial  power spectrum, which 
reduces substructure, and limits the maximum central density of halos, which
eliminates central cusps. 
The relationship between $Q$ and halo observables is estimated.  The primordial
$Q$ may be preserved in the cores of halos and if so leads to a predicted relation, closely
analogous to that in degenerate dwarf stars, between the central
density  and  velocity dispersion.  Classical
polytrope solutions are used to model  the structure of  halos of collisional dark
matter, and to show that self-interactions in halos today are probably not significant
because they   destabilize halo cores
via heat conduction.  Constraints on masses and self-interactions of dark matter 
particles are estimated from halo stability and other considerations. 
\end{abstract}
\newpage

\section{How Cold and How Collisionless is the Dark Matter?}

The successful concordance  of predictions and observations of large scale structure and
microwave anisotropy vindicates many assumptions of standard cosmology,
in particular the hypothesis that the dark matter is composed of
primordial particles which are cold and collisionless\cite{cdm}. 
At the same time,  
   there are hints of discrepancies observed in    the small-scale structure within
galaxy haloes, which we explore as two related but separate issues,
namely the predictions of excessive substructure and sharp central cusps
in dark matter halos.

   The first
``substructure problem'' is that    CDM  predicts excessive relic
substructure\cite{ghigna,klypin}: 
much of the mass of a CDM halo is not smoothly distributed but
is concentrated in many massive sublumps,   like galaxies in 
galaxy clusters. The model predicts that galaxy halos should contain many dwarf
galaxies which  are not seen, and which would disrupt disks even if they are invisible.
The
substructure problem  appears to be  caused by the ``bottom-up'' hierarchical clustering
predicted by CDM power spectra; fluctuations on small scales collapse early and survive as
dense condensations. Its
absence  hints that the small scale power spectrum is filtered  to suppress early collapse
on subgalactic scales. 

 The second ``cusp problem'' is that
CDM  also predicts\cite{dubinski,nfw,moore98,moore99a,moore99b} a universal, monotonic  
increase of  density towards the center of  halos  which is not seen  in close studies of 
dark-matter-dominated galaxies\cite{swaters,swaters2,carignan}
(although the observational issue is far from settled\cite{vandenbosch,dalcanton}).  
The formation of central cusps has been observed for many years in simulations
of collapse of cold matter in a wide variety of circumstances; it may be
thought of 
as   low-entropy material sinking to the center during 
halo formation. Simulations suggest that dynamical ``pre-heating'' of CDM
by hierarchical clustering is
not enough to prevent a cusp from forming--- that some material is always left with a low
entropy and sinks to the center. If this is right, the central structure of halos might 
provide clues to the primordial entropy which are insensitive to complicated
details of nonlinear collapse.

It may be  possible to explain these
discrepancies in a CDM framework\cite{frank}, for example by using various
 baryonic contrivances. It is also possible that the observations can
be interpreted more sympathetically for CDM; we explore this
possibility in more detail in a separate paper\cite{dalcanton}.
  However 
it is also possible
 that the problems with halo structure are 
  giving specific quantitative clues about new properties of the dark matter particles.
By examining halo structure and stability,
in this paper we    make a quantitative assessment
of the effects   of   modifications of the two main properties of
CDM--- the addition of primordial velocity dispersion, and/or the addition of particle
self-interactions. In particular we focus on aspects of halo structure which provide the
cleanest ``laboratories'' for studying   
dark matter properties.  The
ultimate goal of this exercise is  to measure and constrain particle properties from halo
structure.

 Endowing the particles with   non-zero primordial velocity dispersion
 produces two
separate effects:  
a filter in the primordial power spectrum which limits small-scale substructure, 
and a  phase packing  or Liouville limit  which produces halo cores.
Both effects depend on the same   quantity, the 
``phase density'' which we choose to define using the 
most observationally accessible units,
 $Q\equiv
 \rho/\langle v^2\rangle^{3/2}$,
where $\rho$ is the density and $\langle v^2\rangle$ is the velocity dispersion.
The definitions of these quantities depend on whether we are discussing
fine-grained or coarse-grained $Q$.\footnote{For a uniform monatomic ideal thermal gas, 
 $Q$ is related in a straightforward way to
the  usual  thermodynamic entropy; for  $N$
particles,
$S=- kN[\ln (Q)+{\rm constant}].$} 
For collisionless particles, the 
fine-grained $Q$ does not change but the coarse-grained $Q$ can decrease  
as the   sheet occupied by particles folds up in phase space.
The coarse-grained $Q$ can be estimated directly from astronomical 
observations, while the fine-grained $Q$ relates directly to 
microphysics of dark matter particles.
 For particles which decouple 
when still relativistic, the initial microscopic phase density  $Q_0$,
which for nondissipationless collisionless particles is 
the maximum value for all time,  can be related to the particle
mass and type, with little reference to the cosmology. The most familiar 
examples are the standard neutrinos, but  we include in our discussion the 
more general case which yields different numerical factors for
bosons and for particles with a significant chemical potential.

The physics of both  the filtering and the
phase packing in the collisionless case  closely parallels
that of massive neutrinos\cite{gerstein,cowsik},  the standard form
of
 ``hot'' dark matter. Dominant hot dark matter  overdoes both of these
effects--- the filtering scale  is too large 
to agree with observations of galaxy formation (both in   emission
and quasar absorption) and the phase density is too low to agree with 
observations of giant-galaxy halos\cite{tremaine}.  However one can introduce 
new particles with a lower velocity dispersion (``warm'' dark
matter, \cite{bond,bardeen,blumenthal,melott,primack,dodelson}),
which is the option we consider here. 

Although warm
dark matter has most often  been invoked as a solution to fixing  
apparent (and no longer problematic\cite{peacock,cdm}) difficulties
with predictions of the CDM power spectrum for matching galaxy clustering data,
a  
spectrum filtered on smaller scales may also solve  several other classic problems of CDM 
on galactic and subgalactic scales\cite{whiterees,navarrosteinmetz} which are sometimes
attributed to baryonic effects. The  main effect in warm models is that the first nonlinear
objects are   larger and form   later, suppressing substructure and increasing the angular
momentum of galaxies\cite{sommerlarsen}.
This improves the predictions for dwarf galaxy populations\cite{colin},
baryon-to-dark-matter ratio, disk  size and angular momentum, and quiet flows on the scale
of galaxy groups. If the filtering is confined to small scales the predictions are likely
to remain acceptable for Lyman-$\alpha$ absorption during the epoch of
galaxy formation at $z\approx 3$\cite{croft,dave,white}.

Liouville's theorem tells us that dissipationless, collisionless particles can
only decrease their coarse-grained phase density, and 
we conjecture that  halo   cores on small scales 
approximately preserve the primordial
phase density. The
universal character of the phase density allows us to make definite predictions for the
scaling of core density and  core radius with halo velocity dispersion. These relations are 
analogous to those governing nonrelativistic degenerate-dwarf stars: more
tightly bound (i.e. massive) halos should have smaller, denser cores. 
  A  survey\cite{dalcanton} of 
  available
evidence on the phase density of 
 dark matter cores on scales from dwarf spheroidal
galaxies to galaxy clusters shows that the   phase density
needed to create the cores of rotating dwarf  galaxies is much lower than that 
apparently present in dwarf spheroidal
galaxies\cite{aaronson,olszewski,faber,lake,gerhard,ralston,mateo}--- so at  least
one of these populations is not probing primordial phase density. Translating
into masses of neutrino-like relics, the spheroidals prefer masses of about
1 keV (unless the observed stars occupy only a small central portion of an
implausibly large, massive and high-dispersion halo), and the disks prefer about
200 eV. 
   
The larger phase density is also preferred from the point
of view of filtering.  If we take $\Omega\approx 0.3$
(instead of 1  as in most of the original warm scenarios--- which reduces the scale for a
given mass, because it lowers the temperature and therefore the number of the particles),
the filtering scale for 1 keV particles is at about $k=3{\rm Mpc}^{-1}$--- small enough to 
preserve the successful large-scale predictions of CDM but also
large enough to impact the substructure problem.
Galaxy halo substructure therefore favors a primordial phase
density corresponding to collisionless thermal relics with a mass of around
1 keV. 
In this scenario the densest dwarf spheroidals might well preserve the primordial
phase density and in principle could allow a measurement of the particle mass.\footnote{This
raises another unresolved issue: whether the filtering actually prevents systems
as small as dwarf spheroidals from forming at all. The predictions of 
warm dark models are not yet worked out enough to answer this question.}
(Conversely, a  mass as large as 1 keV can only solve the core problem
in disks with 
additional  nonlinear dynamical heating, so that the central matter  no longer remains on 
the lowest adiabat, or with the aid of baryonic effects\cite{frank}.)

 To have the
right mean density and phase density today,  relativistically-decoupling
particles of this phase density  must have
  separated out at least
as early as
 the QCD era, when the number of degrees of freedom was much  
larger than at classical weak decoupling. Their
interactions with normal Standard Model particles must therefore be ``weaker than weak,''
ruling out not only standard neutrinos but many other particle candidates.
The leading CDM particle
candidates, such as   WIMPs and axions,   form in standard scenarios with
much higher  phase densities, although more elaborate
mechanisms are possible to endow these particles with the  velocities
to dilute $Q$.  
We   review briefly some of the available options for making low-$Q$ candidates,
such as particles decaying out of equilibrium.

 A new wrinkle on this story comes if we  endow the particles with 
self-interactions\cite{carlson,delaix,atrio,spergel,mohapatra}.
We  consider a simple parametrized model of particle self-interactions based on
massive intermediate particles of adjustable mass and coupling, and explore the 
constraints on these parameters from halo structure. Self-interactions 
change the filtering of the power spectrum early on, and if they are strong
enough they 
qualititatively change the global structure and stability of halos.

In the interacting case, linear perturbations below
the Jeans scale   oscillate as sound waves instead of damping by free streaming---
analogous to a baryon plasma rather than a neutrino gas.
This effect  introduces a filter which is sharper in $k$ than that from streaming, and 
also on a
scale about ten times smaller than the streaming for the same rms particle velocity--- about
right to reconcile the appropriate filtering
scale with the $Q$ needed for phase-density-limited disk cores. These
self-interactions could be so weak that the particles are   effectively collisionless
today as in standard CDM.

 On the other hand stronger self-interactions have major  effects during the nonlinear
stages of  structure formation and on the structure of galaxy halos\cite{spergel}.  
 We
consider this possibility in some detail, using Lane-Emden polytropes as   fiducial models
for collisional halos. Their structures are close analogs of
degenerate dwarf stars
and we call them ``giant dwarfs''.  We find that these structures are subject
to an instability caused by heat conduction by particle diffusion.\footnote{Degenerate dwarf
stars  are
not subject to this instability because they are supported without
a temperature gradient; the same stabilization could occur in halo cores only if
the dark matter is fermionic and degenerate (e.g., \cite{fuller,shi}). The instability we
discuss here is  essentially what   happens in a thermally-supported star with no nuclear
reactions, except that the conduction is by particle diffusion rather than by radiation.
This effect   may have already been observed
numerically.\cite{hannestad}} Although a
little of this might be interesting (e.g. leading to the formation of central black 
holes\cite{ostriker} or to high-density, dwarf spheroidal galaxies),   typical halos can
only be significantly collisional if they last for a Hubble time;  for this to be the
case,  the particle interactions must be so strong that diffusion is suppressed, which in
turn requires a fluid behavior for all bound dark matter structures. This option is not very
attractive from a phenomenological point of view\cite{delaix,spergel}; for example, dwarf
galaxies or galaxies in clusters tend to sink like rocks instead of orbiting like
satellites, and the collapse of cores occurs most easily in those low-dispersion halos
where we seek to stabilize them.

\section{Particle Properties and Phase Densities}

We  adopt the hypothesis  that some dark matter cores are real and due to dark matter
rather than baryonic physics or observational artifacts. At present this interpretation
is suggested rather than proven by observations.  We
also conjecture   that the   heating  which sets the finite central phase density is
primordial, part of the physics of the  particle creation rather than some byproduct of
hierarchical clustering. At present this is a conjecture suggested rather than proven by
simulations.

 In the clustering
hierarchy, more higher-entropy material is created as time goes on, but  numerical
experiments indicate that this heated material tends to 
end  up in the outer   halo.
This is the basic reason why CDM halos always have central cusps: there
is always a little bit of material which remembers the low primordial entropy
and sinks to the center. The halo center contains the lowest-entropy
material, which we conjecture  is a relic of the original entropy
of the particles---
 or equivalently, their original phase density, which is most directly related to 
measurable properties of halo dynamics. 
We begin by relating the phase density to particle properties in
some simple models.

\subsection{Phase Density of Relativistically-Decoupled Relics}

Consider particles of mass $m$ originating in   equilibrium and
decoupling at
a temperature $T_D>>m $ or chemical potential $\mu>>m $. The original
distribution function is\cite{landau}
\begin{equation}
f(\vec p)=(e^{(E-\mu)/T_D}\pm 1)^{-1}\approx (e^{(p-\mu)/T_D}\pm 1)^{-1}
\end{equation}
with $E^2=p^2+m^2$ and $\pm$ applies to fermions  and bosons
respectively.  The number density and pressure of the particles are\cite{kolb}
\begin{equation}
n ={g\over (2\pi)^3}\int fd^3p
\end{equation}
\begin{equation}
P ={g\over (2\pi)^3}\int {p^2\over 3E} fd^3p
\end{equation}
where $g$ is the number of spin degrees of freedom. Unless stated otherwise,
we adopt units with $\hbar=c=1$.

With adiabatic expansion this distribution
is preserved with  momenta of
particles varying as $p\propto R^{-1}$, so the density
and pressure can be calculated at any subsequent time\cite{peebles}.
For thermal relics $\mu=0$, we can  
derive the density and pressure in the   limit
when the particles have cooled to be nonrelativistic:
\begin{equation}
n={g T_0^3\over (2\pi)^3}\int  {d^3p \over e^p\pm 1}
\end{equation}
\begin{equation}
P={g T_0^5\over (2\pi)^3 3m}\int   {p^2d^3p \over e^p\pm 1}
\end{equation}
where the  pseudo-temperature $T_0=T_D(R_D/R_0)$ records the expansion of
any fluid element relative to its initial size and temperature
at decoupling $R_D,T_D$. 

It is useful to define a ``phase density'' $Q\equiv
 \rho/\langle v^2\rangle^{3/2}$ proportional
to the inverse specific entropy for nonrelativistic matter, 
which  is preserved under adiabatic expansion and contraction. For
nondissipative particles $Q$ cannot increase, although it can decrease due to
shocks (in the collisional case) or coarse-graining (in the collisionless case,
e.g. in
``violent relaxation'' and other forms of dynamical heating.) 
Combining the above expressions for density and pressure
and using $\langle v^2\rangle=3P/nm$, we find
\begin{equation}
Q_X =q_X  g_X m_X^4.
\end{equation}
 The dimensionless coefficient for the thermal case is 
\begin{equation}
q_T={4\pi  \over (2\pi)^3}
{ [\int dp (p^2/e^p\pm 1)]^{5/2}    \over
[\int dp (p^4/e^p\pm 1)]^{3/2}     }=0.0019625,
\end{equation}
where the last equality holds for thermal fermions.
An analogous calculation for the degenerate fermion case 
($T=0, \mu_D>>m_X$) yields the same expression for
$Q $ but with a different coefficient,
\begin{equation}
q_d={4\pi  \over (2\pi)^3}
{ [\int_0^1p^2 dp]^{5/2}    \over
[\int_0^1p^4 dp ]^{3/2}     }=0.036335.
\end{equation}
To translate from $\hbar=c=1$ into more conventional astronomers' units,
\begin{equation}
(100 eV)^4/c^5= 12,808 {(M_\odot/{\rm kpc}^3)\over ({\rm km\ s^{-1}})^{3}}
=12.808 {(M_\odot/{\rm pc}^3)\over (100\ {\rm km\ s^{-1}})^{3}}
\end{equation}
The phase density in this situation depends on the particle properties but not
at all on the cosmology; the decoupling temperature, the current
temperature and density  do not enter. The  numerical factors 
just  depend on whether the particles are thermal or degenerate, bosons or fermions,
which makes the quantity $Q$ a potentially precise tool for measuring particle
properties. 
Many scenarios envision thermal relics so we adopt this as a fiducial reference
in quoting phase densities in $m^4$ units---bearing in mind that the actual
mass may be different in cases such as degnerate sterile neutrinos\cite{shi,fuller},
and that 
for the astrophysical effects discussed below, it is the phase density that
matters.  
For a neutrino-like ($g=2$), thermal relic, 
\begin{equation}
Q_T= 5\times 10^{-4} {(M_\odot/{\rm pc}^3)\over ({\rm km\ s^{-1}})^{3}}
(m_X/1{\rm keV})^4.
\end{equation}

\subsection{Space Density of Thermal Relics}

For  a  standard, relativistically-decoupled
thermal relic, the mean density of the particles can be estimated\cite{kolb}
from   the
number of  particle degrees of freedom at the epoch $T_D$ of decoupling,
$g_{*D}$; the ratio to the critical density is
\begin{equation}
\Omega_X =78.3 h^{-2} [g_{eff}/g_{*D}] (m_X/{\rm 1 keV})
=2.4 h_{70}^{-2} (m_X/{\rm 1 keV}) (g_{eff}/1.5)(g_{*D}/100)^{-1} 
\end{equation} 
where $g_{eff}$  is
the number of effective photon degrees of freedom of the particle  
($=1.5$ for a two-component fermion).
For standard neutrinos which decouple at around 1MeV, $g_{*D}=10.75$.

Current observations suggest that the dark matter density 
$\Omega_X\approx 0.3$ to $0.5$, hence the  mass density for a warm relic
with
$m_X\ge$ 200 eV clearly requires a  much  larger 
$g_{*D}$ than the standard value for neutrino decoupling. Above about 200 MeV,   the 
 activation of the extra gluon and quark degrees of freedom
(24 and 15.75 respectively including $uds$ quarks)
give  $g_{*D}\approx 50$;  activation of heavier modes of the Standard
Model above $\approx 200$GeV produces 
$g_{*d}\approx 100$;  this gives  a reasonable match for
$m_X\approx 200$ eV and  $\Omega_X\le 0.5$, as suggested
by current evidence.  Masses of the order of 1 keV can be accomodated
by somewhat earlier decoupling ($\approx$ TeV) and including many  extra  (e.g.,
supersymmetric or extra-dimensional) degrees of
freedom. Alternatively a  degenerate particle can be introduced via mixing of a sterile
neutrino, combined with a primordial chemical potential adjusted to give the right
density\cite{shi}.
 In any of these cases, the particle must interact
with Standard Model particles much more weakly than normal weak interactions,
which decouple at $\approx 1$ MeV.

Note that   warm dark matter particles   have low 
 densities compared with photons and other species at 1 MeV
 so they do  not strongly affect 
nucleosynthesis. However, their effect is not entirely negligible since they 
are relativistic at early times and add considerably more to the mean total
density in the radiation era than standard CDM particles. 
They add   the
equivalent of $(T_X/T_\nu)^3= 10.75/g_{*d}$
  of an effective extra neutrino species,
which leads to a small increase in the predicted primordial helium abundance  for a
given 
$\eta$. Because the phase density fixes
the mean density at which the particles become relativistic, 
it also fixes this effect on nucleosynthesis (independent of the 
other particle properties, thermal or degenerate etc.) This effect might eventually
become detectable with increasingly precise measurements of cosmic abundances.

\subsection{Decaying WIMPs and Other Particle Candidates}

Thermally decoupled    relics are 
the simplest way to obtain  the required finite
phase density, but they are not the only way.
   Heavier particles can be produced with
 a kinetic temperature higher
than the radiation, accelerated by some  nonthermal process.
Weakly interacting massive particles, including
the favored Lightest Supersymmetric Particles, can
reduce their phase density if they form via out-of-equilibrium
particle decay. A small density of heavy 
unstable particles (X1) can separate out in the standard way, then
 later decay into the present-day (truly stable) dark matter particles (X2).
In a supersymmetric scheme one can imagine for example a gravitino separating
out and decaying into neutralino dark matter.

In the normal Lee-Weinberg scenario for WIMP generation, the 
particle density is in approximate thermal equilibrium until
 $T\approx m_{X}/20$. The particles thin out by annihilation until
their relic density freezes out when the the annihilation rate matches the Hubble
rate, 
$n_{X}\langle \sigma_{ann}v\rangle\approx H$. The density
today is then 
\begin{equation}
\Omega_{X}\approx T_{\gamma 0}^3 H_0^{-2}m_{Planck}^{-3}\langle \sigma_{ann}v\rangle^{-1}
\approx(m_W/100{\rm GeV})^2(m_W/m_X)^2
\end{equation}
where we have used the typical weak annihilation cross section
$\sigma_{ann}\approx \alpha^2m_X^2/m_W^4$ determined by the mass
of the $W$.
The kinetic temperature of the WIMPs freezes out at about
the same time as the abundance, so they are very cold today,
with typical velocities
$v\approx \sqrt{20} T_0/m_X\approx 10^{-14} (m_X/100 GeV)^{-1}$.
This of course endows them with small velocities and an enormous phase density.

A smaller phase density can be produced if these particles
decay at some point into the particles present today. 
If the secondary particles are much lighter than the first,
they can be generated with relativistic velocities at relatively
late times as we require.
Suppose the primary $X1$ particles decay into secondary $X2$ 
particles at a temperature $T_{decay}$. To produce particles with the
velocity $\approx 0.4$ km/sec today (characteristic of a fiducial
200 eV thermal relic phase density), or $v\approx c$ at
$T\approx/300 {\rm  eV}$,
\begin{equation}
m_{X1}\approx m_{X2} T_{decay}/300 {\rm  eV}.
\end{equation}
We also want to get the right density of $X2$ particles. Suppose the 
density of $X1$ is determined by a Lee-Weinberg freezeout,
such that $n_{X1}(T_\gamma\approx m_{X1}/20)\sigma_{ann}v\approx H$.
In order to have $\rho_{X2}\approx \rho_{rad}/600$ at $z_{nr}\approx 10^6$,
$\rho_{X1}\approx \rho_{rad}/600$ at $ T_{decay}$, and then
\begin{equation}
m_{X2}^2T_{decay}\approx{600\times 20m_W^4\over \alpha^2m_{Planck}}
\approx (100{\rm MeV})^3.
\end{equation}
Thus we obtain
\begin{equation}
m_{X1}m_{X2}\approx (30{\rm GeV})^2.
\end{equation}
A simple example would be a more or less standard 50 GeV WIMP primary
which decays at $T_{decay}\approx 1$keV into marginally 
relativistic 20 GeV secondaries. Alternatively the primary could be 
heavier than this and the secondary lighter. Such scenarios have to be crafted
to be consistent with various constraints, such as the long required
lifetime for $X1$ (in the example just given, a week or so) and the decay width of the $Z$
(which must not notice the existence of $X2$); although not compelling, 
they are not all ruled out.\footnote{It is also possible  to 
reduce the scale of filtering of linear perturbations for a given phase density
  by arranging for the decay relatively late, and for the decay products to be
nonrelativistic. This option seems even more contrived and we will not pursue 
it  in detail here.}

The other perennial favorite dark matter candidate is the axion. 
The usual scenario is to produce these by condensation, which if 
homogeneous produces dark matter even colder than the WIMPs--- indeed,
as bosons in a macroscopic coherent state. However, it is natural
to contemplate modifications to this picture where the condensing fields
are not uniform but have topological defects or Goldstone excitations,
produced by the usual Kibble mechanism during symmetry
breaking (e.g. \cite{hoganrees,battye}).  In this case the axions are  produced with
relativistic velocities and could in principle lead to the desired velocity 
dispersion.

\section{Cores from Finite Primordial Phase Density}

We have shown several examples of how particle properties determine 
primordial phase density. Here we explore how the phase density affects 
the central structure of dark matter halos. 

\subsection{Core Radius of an Isothermal Halo}

Consider the evolution of classical dissipationless, collisionless particles in phase
space.  Truly Cold Dark Matter  is formed with zero velocity dispersion occupying a three
dimensional subspace (determined by the Hubble flow $\vec v= H\vec r$) of six dimensional
phase space. Subsequent nonlinear evolution wraps up the phase sheet so that a
coarse-grained average gives a higher entropy and a  lower phase density. In general a small
amount of  cold material remains which naturally sinks to the center of a system. There is
in principle no limit to the central density; the phase sheet can pack an arbitrary number
of phase wraps into a small volume.

By contrast, with   warm dark matter  the initial phase sheet has a finite thickness. The 
particles do not radiate so the phase density  can never exceed this initial value. In the 
nonlinear formation of a halo, the phase sheet evolves as an incompressible fluid in
phase space. The outer
parts of a halo form in the same way as CDM by wraps of the sheet whose thickness is
negligible, but in the central parts the finite thickness of the sheet prevents arbitrarily
close packing--- it reaches a ``phase packing'' limit.   For a given velocity dispersion at
any point in space, the primordial phase density of particles imposes an upper limit on
their density
$\rho$, corresponding  to 
 adiabatic compression.
Thus warm dark matter halos cannot form the singular central cusps
predicted by Cold Dark Matter but instead form cores with a maximum limiting 
  density   at small radius, determined by
the velocity dispersion.

  We estimate   the structure of the halo 
core
 by  conjecturing  that the matter in the central parts of the halo 
lies close to the primordial adiabat defined by $Q$. This
will be a good assumption for cores which form quietly without too much
dynamical heating. Simulations indicate this to be the case in essentially all CDM
 halos, although in principle  it could
be that warm matter typically experiences more additional
dynamical  heating than cold matter,
in which case the   core could be larger.  This question can   be resolved with 
warm simulations, including a reasonable sampling of the particle distribution function
during nonlinear clustering\cite{wadsley}; for the present we derive a rigorous upper
limit  to the core density  for a given  velocity dispersion, and conjecture that this will
be close the actual central structure.   

A useful model for illustration and fitting is
a standard  isothermal sphere model for the halo. The spherical case 
with an isotropic distribution of velocities   maximizes the central density 
compatible with  the phase
density limit. 
The conventional definition of core size in an isothermal
sphere\cite{binney} is the 
``King radius''
\begin{equation}
r_0=\sqrt{9\sigma^2/ 4\pi G \rho_0}
\end{equation}  
where $\sigma$ denotes  the one-dimensional velocity dispersion,
and  $\rho$ denotes the central density.  Making the adiabatic
assumption, $\rho_0= Q (3\sigma^2)^{3/2}$, we find
\begin{equation}
r_0=
\sqrt{9 \sqrt{2}/ 4\pi 3^{3/2}}(QG v_{c\infty})^{-1/2}=
0.44 (QG v_{c\infty})^{-1/2}
\end{equation} 
where $v_{c\infty}=\sqrt{2}\sigma$
denotes the asymptotic circular velocity
of the halo's flat rotation curve.
(Note that aside from numerical factors this is the same mass-radius relation as
a degenerate dwarf star; the galaxy
core is bigger than a Chandrasekhar dwarf of the same specific binding energy by a
factor
$(m_{proton}/m_X)^2$. The collisional case treated below is even closer to a scaled
version of a degenerate dwarf star.)

For the thermal and degenerate phase densities
derived above, 
\begin{equation}
r_{0,thermal}= 5.5 {\rm kpc} (m_X/100{\rm eV})^{-2}
(v_{c\infty}/30{\rm km s^{-1}})^{-1/2}
\end{equation}
\begin{equation}
r_{0,degenerate}= 1.3 {\rm kpc} (m_X/100{\rm eV})^{-2}
(v_{c\infty}/30{\rm km s^{-1}})^{-1/2},
\end{equation}
where we have set $g=2$.
The circular velocity in the central
core displays the harmonic behavior $v_c\propto r$; it reaches half of
its asymptotic  value at a radius $r_{1/2}\approx 0.4 r_0$. 

Instead of fitting an isothermal sphere to an entire rotation curve, in some situations we
might opt to measure the central density directly by fitting the linear inner portion of
a rotation curve if it is well-resolved in the core:
\begin{equation}
v_c/r=\sqrt{4\pi G\rho/3}
= 2.77 G^{1/2}Q^{1/2}v_{c\infty}^{3/2}
= 6.71 {\rm km\ s^{-1} \ kpc^{-1}}(m_X/100{\rm eV})^{2}
(v_{c\infty}/30{\rm km s^{-1}})^{3/2}.
\end{equation}

\subsection{Comparison with galaxy and cluster data}

In a  separate paper\cite{dalcanton} we  review the current 
relevant  data in more detail, including a consideration of interpretive
ambiguities. Here we offer a summary of the situation.

The relationship of core radius or central density with halo velocity
dispersion is a simple prediction of the primordial phase density
hypothesis, which can be in principle be tested on   
a cosmic population of halos. In particular if
phase packing is the explanation of dwarf galaxy cores, the dark matter cores of giant
galaxies and galaxy clusters  are predicted to be much smaller than for dwarfs, 
   unobservably hidden in a central region dominated by
baryons. There is currently at least one well-documented case of a galaxy cluster with a
large core ($\approx 30$kpc) as measured by a lensing fit\cite{tyson}, which  
 cannot be explained at all by phase packing with primordial phase density.
On the other hand   more
representative samples of relaxed  clusters do  not show evidence
of cores\cite{dalcanton,williams}.  

The favorite
 laboratories for finding evidence of dark matter cores are dwarf disk galaxies
which display a central core even at radii where the baryonic contribution
 is negligible\cite{carignan,swaters,swaters2}. Rotation curves allow a direct
estimate of the enclosed density as a function of radius, right out to a fairly
flat portion which allows an estimate of the dark matter velocity dispersion---
all the information we need to estimate a phase density for a core. 
Three of the best-resolved cases\cite{dalcanton} yield estimates 
of $Q\approx   10^{-7}- 10^{-6} (M_\odot/{\rm pc}^3)/ ({\rm km\ s^{-1}})^{3}$.
The sensitive dependence of $Q$ on particle mass means that $m_X$ is 
reasonably well bounded even from just from a handful of such cases; a thermal value of 
$m_X\ge$300 eV does not produce large enough cores to help at all (that is, one
must seek unrelated explanations of the data),  while values $m_X\le$100 eV produce
such large cores that they conflict with observed  rotation curves of normal giant
galaxies\cite{tremaine} and LSB galaxies\cite{pickering}. This is why we
adopt a fiducial reference value of 200 eV for dwarf disk cores.

 Dwarf
spheroidal galaxies do not have gas on 
circular orbits so their dynamics is studied with stellar velocity
dispersions\cite{aaronson,olszewski,faber,mateo}. Here we have an estimate of the mean
density in the volume  encompassed by the stellar test particles, but we do not know the
velocity dispersion of the dark matter halo particles (which may larger than that of  the
stars if the latter occupy only
the harmonic central portion of a large dark matter core) so estimates 
of the phase density are subject  to other assumptions and modeling 
constraints\cite{lake,gerhard}.  If we assume that the stars are not much more
concentrated than the dark matter, we get the largest estimate\footnote{This is the largest
value of the mean phase density of material in the region enclosed by the stellar
velocity tracers;
there is no real observational upper limit for the maximum phase density. Without the
rotation curve information, these systems are consistent with singular isothermal 
spheres or other cuspy profiles for the dark matter} of
the phase density, which in the largest case\cite{dalcanton} is about
$Q\approx 2\times 10^{-4} (M_\odot/{\rm pc}^3)/ ({\rm km\ s^{-1}})^{3}$ corresponding to a
thermal relic of mass
$m_X\approx 800$eV.  The apparent phase densities estimated for dwarf
spheroidals are thus much larger than for dwarf disks, even at the same radius. 
The mass-to-light ratio in the most extreme of these systems is  about 100 in 
solar units, an order of magnitude more than that found for purely baryonic, old
stellar populations in elliptical galaxies\cite{fukugita}, so there is little doubt
that they are dominated by dark matter. The CDM prediction is that there are other, more
weakly bound halos in which gas was unable to cool and form 
stars, and which therefore have an even higher   mass-to-light ratio.

\section{Filtering  of Small-Scale Fluctuations}
The non-zero primordial velocity dispersion  naturally leads to a filtering of the 
primordial power spectrum.
The transfer function of Warm Dark Matter is almost
the same as Cold Dark Matter on large scales, but is
filtered by free-streaming on small scales. The 
characteristic wavenumber for filtering at any time is given
by $k_X= H/\langle v^2\rangle^{1/2}$, the inverse
distance travelled by a particle at the rms velocity in
a Hubble time.  The detailed shape of the transfer function  
depends on the detailed evolution of the Boltzmann equation,
and in particular whether the particles are free-streaming or
collisional.

In the current application,
we are concerned with $H$ during the radiation-dominated era
($z\ge 10^4$), so that $H^2=8\pi G \rho_{rel}/3\propto (1+z)^4 $, where 
$\rho_{rel }$ includes all relativistic degrees of freedom.
For constant $Q$, $\langle v^2\rangle^{1/2}= (\rho_X/Q)^{1/3}\propto
(1+z)$   as long as the $X$ particles
are nonrelativistic. For particles with a small velocity dispersion today,
the    
comoving filtering scale\cite{kolb} is thus approximately independent
of redshift over a considerable interval of redshift (see Figure 1). 
 The ``plateau'' scale is independent of $H_0$:
\begin{equation}
k_{X,comov}= H_0 \Omega_{rel}^{1/2} v_{X0}^{-1}
=0.65\ {\rm Mpc^{-1}} (v_{X0}/1 {\rm km\ s^{-1}})^{-1}
\end{equation}
where 
$\Omega_{rel}= 4.3\times 10^{-5}h^{-2}$ is the density in
relativistic species and  
 $v_{X0}=(Q/\bar\rho_{X0})^{-1/3}$ is the rms velocity of
the particles at their present mean cosmic density
$\bar\rho_{X0}$.
For the thermal   case, in terms of particle
mass, we have
\begin{equation}
v_{X0,thermal}=0.93\ {\rm km\ s^{-1}} h_{70}^{2/3} (m_X/ 100 {\rm
eV})^{-4/3}(\Omega_X/0.3)^{1/3}(g/2)^{-1/3},
\end{equation}
and hence
\begin{equation}
 k_{X,comov}= 15 \ {\rm Mpc^{-1}} h_{70}^{-2/3} (m_X/ 1 {\rm
keV})^{-4/3}(\Omega_X/0.3)^{-1/3}(g/2)^{1/3}.
\end{equation}

In the case of free-streaming, relativistically-decoupled
thermal particles, the transfer function has been computed
precisely\cite{bardeen,sommerlarsen}; the characteristic
wavenumber where the square of the transfer function falls to half 
the CDM value is about  $k_{1/2,stream}\approx  k_{X,comov}/5.5$. 
The simple streaming case only works
for high phase densities $m_X\ge 1$keV, that is, comparable to that
observed in dwarf spheroidals.
For example, to  produce an acceptable number of galaxies at a dwarf galaxy scale
without invoking disruption,
Press-Schechter theory\cite{kamionkowski} implies a spectral
cutoff  at about $k=3 h_{70} {\rm Mpc}^{-1}$, 
requiring a thermal relic mass of about 1100 eV. Hydrodynamic 
simulations show that the same cutoff scale
preserves the  large scale success of CDM and probably improves the 
CDM situation on galaxy scales in ways mentioned previously\cite{sommerlarsen}. 
 Although the typical uncertainty
on the phenomenologically best filtering scale is at least a factor of two, 
it is clear that the smallest phase density  compatible with standard streaming filtering
is    too large to have
a direct impact on the core problem in dwarf disk galaxies.

On the other hand the discrepancy is only   a factor of a few in mass,
less than an order of magnitude in linear damping scale. We have already mentioned two
  modifications which could reconcile these scales. It could be that  warm
models turn out to
 be sometimes more effective at producing smooth cores
than we have guessed from the minimal phase-packing  constraint,  due to more efficient
dynamical heating than   CDM; this would produce   a
  nonlinear amplifier of the primordial velocities, probably with a large 
variation depending on dynamical history  (an especially good option if cores turn out to
be common in galaxy clusters.) Another
possibility is that the primordial velocities are introduced relatively late
(nonrelativistically) by particle decay.

Still another possibility is  a different relationship of
$k_{1/2}$ and $k_X$ from the standard collisionless streaming behavior. For example, if the
particles are self-interacting, then  the free streaming is suppressed and the relevant
scale is the standard Jeans scale dividing  growing behavior from
acoustic oscillations, $4\pi G\rho_{total}-k_J^2c_S^2=0$. This comes out to
$k_J=\sqrt{3}H/c_S=\sqrt{27/5}k_X$, 13 times  shorter
than  $k_{1/2,stream}$ at a fixed phase density. (An intuitive view of
the this numerical factor is that
during the long period when $k_X$ is flat, streaming particles continue
to move and damp on larger scales, whereas the comoving Jeans scale just remains fixed,
sharply dividing oscillating from growing behavior.)
The acoustic case is similar to the behavior of fluctuations in high-density,
baryon-dominated models, which have a sharp cutoff at the Jeans
scale\cite{peacock}.
We conclude  that some particle self-interactions may be 
desirable to reconcile  the scale of the  transfer function of primordial 
perturbations with the phase packing effect on disk cores.

\section{Collisional Dark Matter}

 We now turn to    the case  where the 
 dark matter particles are not collisionless, but scatter off of each other
via  a new intermediate force. Self-interactions of dark matter have
been motivated from both an astrophysical and a particle physics point of
view\cite{carlson,delaix,atrio,spergel,mohapatra}. Our goal here is
again to relate the  properties of the new particles to the potentially observable
properties of  dark matter halos. In addition to the single parameter $Q$ considered for
the  collisionless case, we can use halo properties to constrain
fundamental parameters of the particles---  the masses of the dark matter particles  
and intermediate bosons carrying the interactions, as well as a coupling constant. 

 Such
self-interactions lead  to modifications in several of the previous arguments.
 As we have seen, self-interactions can have
observable  effects via the transfer function even if they are negligible today.  
Stronger self-interactions   also affect the structure and stability of halos;
collisional matter has a fluid character leading to  
equilibrium states of self-gravitating halos much like those of stars. 
These systems are quite different from collisionless systems. Although entropy must 
increase outwards for stability against convection (which naturally happens
due to shocks in the hierarchy), it cannot
increase too rapidly and remain hydrostatically stable; in particular, stable solutions
have a minimum nonneglible 
temperature gradient, and the
isothermal case is no longer a stable static solution as it is for collisionless matter.
Since collisional matter   conducts heat between fluid elements, these solutions
are all unstable on some timescale. 

\subsection{Particles and Interactions}
We now apply several simple physical arguments to constrain properties of 
the dark matter candidate and its interactions. Some of these have been considered
previously\cite{spergel}.  The most important constraints are summarized in 
figure 2.

Suppose that the dark matter 
 $X$ particles with mass $m_X$, which may be either
fermions or bosons, interact
via massive bosons $Y$ whose mass $m_Y$ determines the
range of the interactions, and a coupling constant
$e$. These may be considered analogous to strong interaction
scatterings 
where we regard pions as Yukawa scalar intermediates, or electroweak
interactions with $W,Z$ as vector intermediates.
The interactions must be elastic scatterings to avoid 
a net energy loss, although ``dissipative'' three-body
encounters are permitted as long as the energy does not leave
the $XY$ subsystem nor travel far in space. 
 For most purposes even
the sign of the interaction does not matter--- it may be attractive  or
repulsive, as happens with vectors and like charges.
The  $Y$ particles at tree level interact only with
$X$, although the $X$ may (as is usual with dark matter candidates) be allowed some
much weaker interactions with ordinary matter.
In this model the collision cross section
for strong scattering is about
\begin{equation}
\sigma\approx m_Y^{-2} {\rm min}
\left[e^4\left({m_Y\over m_X v^2}\right)^2,
\ e^4 \left({m_X\over m_Y}\right)^2,
\ 1\right]
\end{equation}
where the first case is coupling-limited (and depends
on the particle velocity and coupling strength, like electromagnetic
scattering of electrons),
the second case holds for $m_Y>m_X$ (like neutrino
neutral-current interactions)  and the third is the
range-limited, strong interaction limit (like neutron scattering).

There are several  simple constraints on the particle masses.
If the dark matter is collisional, 
the rate of net annihilations of $X$ must be highly
suppressed compared to the scattering rate, or the 
mass of the halo would quickly radiate away as $Y$ particles.
Either there is a primordial asymmetry
(so the number of 
$\bar X$ is negligible), or 
\begin{equation}
m_Y>2m_X,
\end{equation}
 suppressing
what would otherwise be a  rapid channel for $X$ to annihilate and
radiate $Y$. (Recall that in this model, there is no direct route to annihilate
into anything else).   In any case
the
$Y$ must not be too light or the typical inelastic collisions will radiate them;
for particles with relative velocities $v\approx 10^{-3}$ typical of dark matter in 
galaxies, we must have 
\begin{equation}
m_Y> m_X v^{2}\approx 10^{-6} m_X,
\end{equation}
so that the energy of collisions is typically insufficient to 
create a real $Y$.
In addition, if attractive, the range of the interactions must be less than
the ``Bohr radius'' for these interactions, requiring
\begin{equation}
m_Y>e^2m_X,
\end{equation}
 in order not to form bound ``atoms''.
The close analogy with $Y$ is  the pion, which is just
light enough to allow a bound state of deuterium.
Bound states would be a disaster since they would behave like
nuclear reactions in stars. 
Such states would  add an internal source of energy
in the halos, creating winds or other energy flows which  would  
unbind large amounts of    matter. All of these constraints eliminate the upper
left region of figure 2, with details depending on the coupling strength and
halo velocity. 

\subsection{Parameters for Collisional Behavior}

The   properties of interacting particles define a characteristic column density,
 $m_X/\sigma$; a slab of $X$ at this column is one mean free path thick.
This is the quantity  that specifies the degree of  collisional or  collisionless behavior
of a system. In order to connect the halo astrophysics with dark matter properties
we  convert from units with $\hbar=c=1$:
\begin{equation}
(1\ {\rm GeV})^3= 4.6\times 10^3\ {\rm g\ cm^{-2}}
=2.2\times 10^7\ {\rm M_\odot\ pc^{-2}}
\end{equation}
For comparison, the average mass column density
within radius  $r_{kpc}{\rm kpc}$ for a halo with a circular velocity
$v_{30}\times 30{\rm km\ sec^{-1}}$
     is 
\begin{equation}
\Sigma_h= {v^2\over \pi Gr}= 0.014\ v_{30}^2 r_{kpc}^{-1}
{\rm g\ cm^{-2}}=(15{\rm MeV})^3v_{30}^2 r_{kpc}^{-1}.
\end{equation}
A halo therefore enters the strongly-collisional regime---
qualitatively different from classical CDM--- if
\begin{equation}
m_Y^4e^{-4}(15{\rm MeV})^{-3}v_{30}^{-2}  r_{kpc} < m_X<\min[(15{\rm MeV})^3v_{30}^2
r_{kpc}^{-1}m_Y^{-2},
 15{\rm MeV}(e/v)^{4/3}(v_{30}^2 r_{kpc}^{-1})^{1/3}].
\end{equation}
This criterion is shown in figure 2 as the right boundary of the ``unstable cores''
region; indeed  this marginally-collisional case maximizes the rate of thermal
conduction instability, as discussed below.

We also compute
the criterion for non-streaming behavior in the early universe--- the amount of
self-interaction needed  to affect the transfer function as discussed above.
It  is
significantly less than that required for collisional behavior today:
\begin{equation}
{\sigma\over m_X}
\approx H(t_{eq})/n_X(t_{eq})v_X(t_eq)m_X
=\Sigma_0^{-1}\Omega_{rel}^{5/2}\Omega_X^1 v_X(t_0),
\end{equation}
where $eq$   refers to the epoch of equal densities in dark matter and relativistic
species, and
\begin{equation}
\Sigma_0\equiv c\rho_{crit}/H_0=0.1213 h_{70}^{-1} {\rm g\ cm^{-2}   }
\end{equation}
is the characteristic cosmic column density today.
Using the units conversion above we have
\begin{equation}
{\sigma\over m_X}
\approx(600 {\rm MeV})^{-3}(v_{X0}/1 {\rm km \ s^{-1}})^{-1}
(\Omega_X/0.3)^{2}h_{70}^4,
\end{equation}
corresponding to a mass column for one expected  scattering
of $2\times 10^4 {\rm g\
cm^{-2}}$. 
Particles scattering off of each other more strongly than this  no longer have streaming
behavior at high redshift but support acoustic  oscillations, much like baryons but with
only their own pressure (that is, without the interaction with radiation pressure and
without decoupling from it).  We should bear  in mind that a  somewhat larger cross section
is needed to avoid diffusive (``Silk'') damping,  but even  at this level of interaction
the scale of damping is is significantly reduced from the streaming case. 
This criterion is shown in figure 2 as the right boundary of the
``Jeans''
region (although some acoustic behavior before $t_{eq}$ occurs even to the right of this).

\subsection{Polytropes}

The equilibrium configurations of collisional dark matter correspond to those
of classical self-gravitating fluids.  The simplest cases to consider and
general enough for our  level of precision are classical polytrope solutions---
 stable configurations of a classical,
self-gravitating, ideal gas with a polytropic equation  of state.\cite{zeldovich}
In the absence of shocks or conduction,   the pressure
and density of a fluid element obey an equation of state $p=K_1\rho^{\gamma_1}$.
 For an adiabatic, classical, nonrelativistic, monatomic gas,
or for nonrelativistic degenerate particles,
the adiabatic index
$\gamma_1=5/3$ and different values of $K_1$ correspond to different entropy.
If   the 
entropy varies radially as a power-law,
 equilibrium 
self-gravitating  configurations  
are given by
 classical Lane-Emden polytrope solutions. The radial variation of  pressure and density
obey
$p(r)=K_2\rho^{\gamma_2}(r)$;  the second index
$\gamma_2$  tracks the radial variation  between different fluid elements
in some particular configuration (that is, including variations
in entropy).
For gas on the same adiabat everywhere, $\gamma_1=\gamma_2$;
for the case of nonrelativistic degenerate or adiabatic matter,   $\gamma_1=\gamma_2=5/3$
 applies and is a good model of degenerate   dwarfs. 
If 
the entropy is increasing with radius, as would be expected
if assembled in a cosmological hierarchy, then $\gamma_2< 
\gamma_1$, conferring stability against convection.

The character of the solutions is well known\cite{zeldovich}.
 As long as $\gamma_2> 6/5$ the
halo structure is  like a star, with a flat-density core in the center, falling off in the
outer parts to vanishing  density at a  boundary. If it is rotating, the 
structure is similar but rotationally flattened.  
These solutions describe approximately the structure of stars,
especially degenerate dwarfs, and 
halos of highly collisional dark 
matter.\footnote{It is worth commenting on some differences and similarities with 
collisionless  halos with finite phase density material. The polytrope solutions
are for collisional matter with an isotropic pressure and local balance of pressure
gradient and gravity. Collisionless particles can fill phase space more sparsely, but this
just means that at a given mass density they must  have a larger maximum velocity;
the collisional solution  saturates the phase density limit and
 has  the largest
mass density for a given coarse-grained phase density.  In this sense,
once one is solving the cusp problem with finite phase density,
nothing further is gained by making the particles collisional.
Collisionless particles allow  anisotropy in the momentum
 distribution function, and therefore a wider
range of ellipsoidal figures, but cannot pack into tighter cores. 
For the same reason, the inner phase-density-limited
 core is expected to be close to
spherical except for rotational support, whether the particles are collisional or not.
The phase space is fully occupied and therefore the velocity
distribution is close  to isotropic wherever the local entropy approaches 
the primordial value.}
If $\gamma_2<6/5$ 
(and in particular for the isothermal 
case $\gamma_2=1$) there is a dynamical instability   
and no stable solution; the system runs away on a gravitational
timescale, with the center collapsing and the outer layers 
blowing off.

\subsection{Giant Dwarfs}
 At zero entropy the equilibrium configuration is
the exactly soluble $\gamma=5/3$ polytrope, which we adopt as
an illustrative example. That is, we model a dwarf galaxy core
as a degenerate dwarf star, the only difference being a  
particle mass much smaller than a proton
 allowing a   halo mass much bigger than a star. For total mass $M$
and radius
$R$,   the Lane-Emden solution gives a central pressure   
   $p_c=0.770
GM^2/R^4$  and a central density   $\rho_c=5.99\bar\rho=1.43 M/R^3$.
Using the above relation for the equation of state we 
obtain the standard degenerate dwarf solution, which has
\begin{equation}
R= 4.5 m_X^{-8/3} M^{-1/3} m_{Planck}^{2}
=0.98{\rm kpc}\left({m_X\over 100{\rm eV}}\right)^{-8/3}
\left({M\over 10^{10}M_\odot}\right)^{-1/3},
\end{equation}
where $m_{Planck}=\sqrt{\hbar c/G}$ and $M_\odot=9.48\times 10^{37}m_{Planck}$.
 This ``giant dwarf'' configuration is stable even at zero temperature up to
the Chandrasekhar limit for $X$ particles.\footnote{Defined
analogously to the Chandrasekhar limit for standard dwarfs (with $Z=A$ because there is just
one kind of particle providing both mass and pressure, similar to a neutron star),
\begin{equation}
M_{CX}= 3.15 {m_{Planck}^3\over m_X^2}
=4.95\times 10^{14}M_\odot(m_X/100{\rm eV})^{-2}.
\end{equation} }

Since the mass is not directly observable, it is more
useful to consider the velocity of a circular orbit at the surface,
$v_c=(GM/R)^{1/2}$.
We then obtain the relation for a degenerate system,
\begin{equation}
m_X= 4.5^{3/8} v_c^{-1/4} (r_cm_{Planck})^{-1/2} m_{Planck},
\end{equation}
or in  more conventional astrophysical units,  
\begin{equation}
m_X= 87 eV \ (v_c/30{\rm km/s})^{-1/4} (r_c/1 kpc)^{-1/2}.  
\end{equation}
Note that as in the collisionless case, no cosmological assumptions or 
parameters have  entered into this expression.

For any  adiabatic nonrelativistic matter the solution is similar.
The absolute
scale  of the giant dwarf, determined by $K_2$,
  is fixed by the phase density $Q$.  In general there is a range of entropy but 
once again the 
 the lowest-entropy  material (which is densest at a given pressure) 
sinks to  the center  of a halo
 and forms an approximately adiabatic core.  The rest of the halo forms
a thermally-supported atmosphere
above it. Once again cores are the places to look for signs of a primordial
ceiling to phase density. 
However, as we see below  the behavior changes if  conduction or radiation are not
negligible. As we know, a thermally supported star which conducts heat
and has no nuclear or other source of energy is 
unstable.

\subsection{Heat Conduction Time and Halo Stability}
If the collision rates are not very high we must consider
heat and momentum transport between fluid elements
by particle diffusion. The most
serious consideration for radial stability is the transport of heat.
  In all
stable thermally supported solutions the 
dense inner parts are  hotter;  if  
conduction is allowed, heat is transported outwards.  The entropy
of the central material decreases,  
the interior is compressed to higher density and the outer
layers spread to infinity, a manifestation of the gravothermal catastrophe.
With conduction   the inner gas falls in and
the outer gas drifts 
out on a diffusion timescale, attempting to approach
 a  singular isothermal sphere. 

Consider the scenario\cite{spergel} where  the dark matter cross
section is small enough to remain essentially noninteracting on large scales,
preserving the successes of CDM structure formation simulations,
but large enough to become collisional 
 in the dense central regions of galaxies.
Although this scenario was introduced to help solve the 
cusp problem, we will see that the conductive instability
  makes matters worse. If stable cores
are to last for  a Hubble  time, the dark matter halos must either be effectively
collisionless (standard dark matter),
or very  strongly interacting, so that the inevitable conduction is
slow (or made of degenerate fermions so there is no temperature gradient.)

Elementary kinetic theory\cite{landau}  yields an estimate
for the  the conduction
of heat by particle diffusion;
the ratio of energy flux to temperature
gradient is the classical conduction coefficient $\kappa\approx
\sigma^{-1}\sqrt{T/m}$. Assuming a halo in approximate virial
equilibrium and profile $v(r)$, this yields a timescale for
heat conduction,
\begin{equation}
t_{cond}\approx   {v \sigma\over 2 Gm_X} {-d\log r\over   d \log v}
\end{equation}
where    $v$ is the typical particle velocity
(which is about the virial velocity of the halo independent of the 
mass of the particles $m_X$). The first factor is essentially
the time it takes a particle to random walk a distance $r$,
$t_{diffuse}\approx {r^2 n \sigma/v}$.
 The last factor
characterizes the temperature and entropy gradient; dynamical stability
prevents it from being very large, and in most
of the matter it typically takes a value not much larger than unity.\footnote{The conductive
  destabilization probably  happens
faster than Spergel and Steinhardt estimated.
They used the Spitzer formula describing core collapse in
globular clusters, which takes about 300 times longer
than the two-particle relaxation time. However, the large factor
arises because  in the globular cluster case the relaxation is
entirely gravitational and is dominated by very long-range
interactions with distant stars. In the present situation  the interactions are
strong and short-range, leading to significant exchange of
both energy and momemtum in each scattering.  The transport of
heat takes place on the same timescale as the diffusion of particles, with numerical
factors of  the order of unity as in standard solutions of the Boltzmann
equation for gases.}

A halo with conduction therefore forms a kind of cooling flow, with the core collapsing and
the  envelope expanding. If it is hydrostatically quasi-stable (that is, if the core
collapse is slow and  regulated by the particle diffusion),
we can use the Lane-Emden solutions to set bounds on the numerical factor
${d\log r/   d \log v}$ governing the instability. 
The equation of state tells us that
 $v\propto \rho^{1/2n_2}$ where $n_2=(\gamma_2 -1)^{-1}$. The largest value of $n_2$
which corresponds to a quasi-stable solution is $n_2=5$. The density
profile is steeper than isothermal ($n_2=\infty$), $\rho\propto r^{-2}$;
therefore $|{d\log r/   d \log v}|\le n_2 \le 5$. In the rough estimates here
we set these factors to unity.\footnote{Another interesting limit is that
of small but nonzero self-interactions.   
The halo is  essentially collisionless, but  occasional scatterings still take
place. The collisionless isothermal sphere, singular or not, is then
an approximate solution, but  still subject to a slow secular instability from
heat conduction.
It is also possible to set up situations where halos are evaporated by a hot
external environment, heated from outside by collapse of the cosmic web.}

Conduction can be suppressed if the scattering is very frequent.
For nondegenerate $X$,   stable
cores require that 
 the conduction time exceeds the Hubble time $H_0^{-1}$.
For stability over a Hubble time, the  column density of a 
halo with velocity $v$ must exceed ${m_X/ \sigma}=  Hv/G$; therefore
the particles must satisfy
\begin{equation}
{m_X\over \sigma}\ge
1.0\times 10^{-4} {\rm g\ cm^{-2}} h_{70}  v_{stable,30}.
\end{equation}
where $v_{stable,30}\times 30{\rm km s^{-1}}$   denotes the  
velocity  in the lowest-velocity stable halo.
Perhaps surprisingly, the mass and radius of the halo
do not enter explicitly. 

This condition constrains the
particles to be  highly interactive.
Galaxy halos   have slow conduction compared to $H$ only 
above a critical velocity dispersion 
$v_{crit}\approx (G/H)(m_X/\sigma)$.
Halos   below this threshold
should have collapsed cores, and  above the threshold the
core radius/mass relation  is   determined as before by
 the giant dwarf sequence for the the particle's phase density.
The existence of stable bound  30 km/s halos of highly-collisional dark matter
requires
\begin{equation}
{m_X\over \sigma}\le 
(2.8 {\rm MeV})^3 h_{70}  v_{stable,30},
\end{equation}
shown in figure 2 as the right boundary of the ``fluid'' region. 

The ``thickness'' of a  halo with velocity $v_{30}\times 30{\rm km s^{-1}}$,
in units
of particle pathlengths, is
\begin{equation}
{\Sigma_h\sigma\over m_X}\approx
10^2 v_{30}^2r_{kpc}^{-1}h_{70}^{-1}v_{stable,30}^{-1}
\end{equation}
so it is clear that all dark-matter-dominated structures,
from small galaxies to galaxy clusters
($v_{30}\approx 1-  30$, $r_{kpc}\approx 0.1-  1000$), are highly 
collisional and their dark matter behaves as a fluid.
 Even for very diffuse 
matter at the mean cosmic density ($\Omega_X=0.3$), the 
particle mean free path is at most $12 v_{core,30}h_{70}^{-1}$Mpc,
about the same as the scale of nonlinear clustering,
so   all bound dark matter structures act like fluids.

Are other data consistent with the idea 
that essentially all dark matter acts like a fluid? This option
has been considered previously\cite{delaix} and while it is perhaps
not definitively ruled out, it is not phenomenogically 
compelling. Serious problems arise for example
from satellite galaxies which are thought\cite{johnston} to have had several 
orbits without stopping and sinking as they would in a fluid,
or from galaxies in clusters, at least some of which appear (from lens
reconstruction mass maps) to have retained some of their dark matter halos.  An
intriguing possibility is that a small collision rate might contribute to 
enough instability to feed the formation of black holes\cite{ostriker}. However
the rate of the instability is greatest in the lowest mass, lowest density 
galaxy cores, a trend not conspicuous in the demography  of central black holes of
galaxies\cite{magorrian}.\footnote{We have to take note of another possibility: perhaps the
dwarf spheroidals, which  have the lowest velocity dispersions of all galaxies and
are also the densest,  have already collapsed by heat conduction. In this  way
we could     use phase packing to give the cores of
the dwarf disk galaxies and still explain  why the dwarf spheroidals 
have such a large phase density. Note that this scheme also gives the right
filtering scale since the particles are collisional at early times.  The 
dwarf spheroidals  need
not of course collapse all the way to black holes,
but they may well have singular dark matter profiles.} 

We conclude that dark matter self-interactions are likely to be negligible in galaxy halos,
and that this places significant constraints on the particles.
Figure 2 summarizes the constraints on the parameters $m_X,m_Y$ of
 this interacting-particle model
from  the various constraints considered here.
\section{Conclusions}
We have found that some halos might preserve in their inner structure
  observable  clues to new dark matter physics, and   that  indeed
some current observations already hint that the dark matter might be warm rather
than cold. 
We conclude with a summary:

1. Halo cores can be created by a ``phase-packing limit''
depending on finite  initial phase density. They
may provide a direct probe of
  primordial velocity dispersion in dissipationless dark
matter. 

2. For relativistically-decoupled thermal relics, the phase density depends on 
the particle mass and spin but not on cosmological parameters.

3. Rotation curves in a few dwarf disk  galaxies indicate cores with 
a phase density  corresponding to that  of a 
200 eV thermal relic or  an rms velocity of about 0.4
km/sec at the current cosmic mean density. Velocity dispersions in 
dwarf spheroidal galaxies indicate a  higher phase density, corresponding
to a thermal relic mass of about 1 keV. At most one of these populations
can be tracing the primordial phase density.

4. Thermal relics in this mass range can match the mean cosmic density with a plausible
superweak decoupling from Standard Model particles before the QCD epoch. 

5. Other very different particles are consistent with the 
halo data, provided they have the about the same mean density and phase density.
Examples include    WIMPs from particle decay and axions from defect decay.

6. Cores due to phase packing limited by primoridial $Q_0$ predict 
  a
universal relation between core radius and halo velocity dispersion.
 The relation is not found in a straightforward interpretation of the data.

7. Primordial velocity dispersion  also suppresses halo substructure
(and solves some other
difficulties with CDM)  by filtering  
primordial adiabatic perturbations. Estimates based on luminosity functions prefer
filtering  on a scale of about
$k\approx 3 {\rm Mpc}^{-1}$; 
for collisionless particles, this scale
corresponds to a filter caused by
streaming of  about a 1keV thermal relic.

8. Weak self-interactions change  from streaming to acoustic
behavior, reducing  the damping  scale and sharpening the filter.

9. Stronger self-interactions
destabilize halos by thermal conduction, making the cusp 
problem worse (unless
they are very strong--- too strong for satellite-galaxy  kinematics---
 or particles are degenerate, eliminating the central temperature gradient).

10. A simulation which samples a warm distribution function reasonably well is
strongly motivated, to determine whether primordial $Q$ is preserved in the centers
of halos, or whether nonlinear effects can amplify dynamical heating in 
such models to explain cores on all scales.


\acknowledgements
We are grateful for useful discussions  of these issues with
F. van den Bosch,   A. Dolgov,
G. Fuller, B. Moore, J. Navarro,  T. Quinn, J. Stadel,
J. Wadsley, and S. White.
JD gratefully acknowledges the hospitality of the Institute
for Theoretical Physics at UC Santa Barbara, which is supported
in part by the National Science Foundation under Grant No.\ PHY94-07194.
JD was partially supported by NSF Grant AST-990862. CJH thanks
the Max-Planck-Institute f\"ur
Astrophysik,     
  the Isaac Newton Institute for Mathematical
Sciences  and the Ettore Majorana Centre for Scientific
Culture for hospitality.
His work was supported at the University of Washington
by NSF and NASA, and at the Max-Planck-Institute f\"ur
Astrophysik by a Humboldt Research Award.

\begin{figure}[htbp]
\caption{Characteristic masses and velocities as
 a function of inverse scale factor
$(1+z)$, for a cosmological model with $\Omega_X=0.3$,
$\Lambda=0.7$. 
Mass and velocity are plotted in units with $H_0=\bar\rho=c=1$,
or $M=0.3\rho_{crit} c^3H_0^{-3}=1.56\times 10^{21}h_{70}M_\odot$.
The total rest mass of dark matter 
in a volume $H^{-3}$ is denoted by $Hx$; total mass-energy of all forms
in the same volume is denoted by $H$.  
Characteristic rms velocities and streaming masses (rest mass of $X$ in a volume
$k_X^{-3}$) are also shown,
for dark matter with three different phase densities.  The cases plotted correspond to
relativistically-decoupled thermal relics decoupling at three different effective degrees
of freedom, corresponding to 1, 8, and 80 times that for a single standard massive
neutrino--- ``hot'', ``warm'', and ``cool''. (For $h=0.7$, the corresponding masses are 13,
108, and 1076 eV respectively, and the  rms velocities at the present epoch are  $1.3\times
10^{-5}$,
$7.9\times 10^{-7}$, and$3.6\times 10^{-8}$, respectively).
Note the
 long flat period with nearly constant comoving $k_X$ for the cool  particles, during
the period when  the universe
is radiation-dominated but $X$ is nonrelativistic.
The difference between
streaming and collisional behavior during this period has a significant
effect on the scale of filtering in the transfer function, with a sharper 
cutoff and a smaller scale (for fixed $k_X$) in the collisional Jeans limit.}
\end{figure}

\begin{figure}[htbp]
\caption{A sketch of the principal  
constraints from halo structure arguments  on the masses of collisional dark matter
particle $X$ and   particle mediating its self-interactions, $Y$.
This plot assumes a coupling constant $e=0.1$. 
The rightmost   region is indistinguishable from standard collisionless
CDM. The region labled ``Jeans'' 
is essentially collisionless today, but collisional before $t_{eq}$ and consistent
with other constraints; in this regime the particles are no longer
free-streaming, and the filtering scale and the
shape  of the transfer function are significantly modified by self-interactions. 
Somewhat stronger interactions lead to a conductive instability in halos; 
the ``unstable cores'' constraint is ruled out if we require stability down to 
halo velocities of 30 ${\rm km \ s^{-1}}$. 
The leftmost region (``fluid'') produces  halos which are 
so collisional they are stable against conduction for a Hubble time, 
but is probably ruled out by the unusual fluid-dynamical behavior this
would cause in the trajectories of satellite galaxies and galaxies in clusters.
The upper constraint comes from suppression of the annihilation channel (by the inability
to radiate $Y$); if
this  does not apply (that is, if there there no $\bar X$ around) then   parallel,
somewhat higher constraints come  from suppressing dissipation by $Y$ radiation, or
from the prohibition against bound $X$ atoms.
  The  bottom
constraint corresponds to a phase-packing limit for giant galaxies; this last 
constraint on mass applies for relativistically-decoupled light relics only, and is
ten times higher if we use the limit from dwarf spheroidals.}
\end{figure}



\begin{thebibliography}{}
\bibitem{aaronson}
Aaronson, M. 1983, ApJ 266, L11
\bibitem{atrio}
Atrio-Barandela, F., and Davidson, S. 1997, Phys. Rev. D 55, 5886  
\bibitem{bardeen}
Bardeen, J. et al. 1986, Astrophys. J. 304, 15
\bibitem{battye}
Battye, R. A. and Shellard, E. P.S. 1994, Nucl. Phys. B423, 260
\bibitem{binney}
Binney, J.  and Tremaine, S. 1987, {\it Galactic Dynamics}, (Princeton:
Princeton University Press)
\bibitem{blumenthal}
Blumenthal, G.
  Pagels, H. and   Primack, J. R.  1982,   Nature, 299, 37
\bibitem{bond}
Bond, J. R., Efstathiou, G., and Silk, J. 1980, Phys. Rev. Lett. 45, 1980
\bibitem{frank}
van den Bosch, F. et al.  1999, Astron. J., submitted (astro-ph/9911372)
\bibitem{bullock}
Bullock, J. S. et al. 2000, MNRAS in press, astro-ph/9908159
\bibitem{carignan}
Carignan, C. 1985, Astrophys. J.S 58, 107
\bibitem{carlson}
Carlson, E. D., Machacek, M. E. and Hall, L. J. 1992, Astrophys. J.  398, 43
\bibitem{colin}
 Colin, P.,   Avila-Reese, V.,  
and
 Valenzuela, O. 2000,
submitted to the Astrophysical Journal, astro- ph/ 0004115  
\bibitem{colombi}
Colombi, S., Dodelson, S., and Widrow, L. M.1996, Ap. J. 458, 1
\bibitem{cowsik}
Cowsik, R., and McClelland, J. 1972, Phys. Rev. Lett. 29, 669
\bibitem{croft}
Croft, R. A. C. et al. 1999, Astrophys. J. 520, 1
\bibitem{dalcanton}
Dalcanton, J. and Hogan, C. J. 2000, submitted to Astrophys. J., astro-ph/0004381
\bibitem{dave}
Dave, R., Hernquist, L, Katz, N., Weinberg, D. H. 1999, Astrophys. J. 511, 521 
\bibitem{delaix}
de Laix, A., Scherrer, R. J., and Schaefer, Astrophys. J. 452, 495
\bibitem{dodelson}
Dodelson, S. and Widrow, L. M. 1994, Phys. Rev. Lett. 72, 17
\bibitem{dubinski}
Dubinski, J. and Carlberg, R. 1991, Astrophys. J. 378, 496
\bibitem{faber}
Faber, S. and Lin, D. 1983, Astrophys. J. 266, L21
\bibitem{fukugita}
Fukugita, M., Hogan, C. J., and Peebles, P. J. E. 1998,
Astrophys. J. 503, 518
\bibitem{fuller}
Fuller, G., 1999,  preprint
\bibitem{gerhard}
Gerhard, O. E. and Spergel, D. N. 1992, Astrophys. J. 389, L9
\bibitem{gerstein}
Gerstein, S. S. and Zeldovich, Ya. B. 1966, ZhETF Pis'ma Red. 4,174
\bibitem{ghigna}
Ghigna, S., Moore, B., Governato, F., Lake, G., Quinn, T., and Stadel, J.
1999, Astrophys. J. submitted (astro-ph/9910166)
\bibitem{hannestad}
Hannestad, S., 1999, astro-ph/9912558
\bibitem{hoganrees}
Hogan, C. J. and Rees, M. J. 1988, Phys. Lett. B205, 28
\bibitem{hogangoldstone}
Hogan, C. J. 1995, Phys. Rev. Lett. 74, 3105
\bibitem{johnston}
Johnston, K. V., Majewski, S. R., Siegel, M. H., Reid, I. N.,
and Kunkel, W. E. 1999, Astron. J. 118, 1719
\bibitem{kamionkowski}
Kamionkowski, M. and Liddle, A. R. 1999, astro-ph/9911103
\bibitem{klypin}
Klypin, A., Kravtsov, A. V., Valenzuela, O., and Prada, F. 1999, Astrophys. J., 523, 32
\bibitem{kolb}
Kolb, E. W. and Turner, M. S. 1990, {\it  The Early Universe},
(Redwood City: Addison-Wesley)
\bibitem{lake}
Lake, G. 1990, Astrophys. J. 356, L43
\bibitem{landau}
Landau, L. D.  and Lifschitz, E. M. 1958, {\it Statistical Physics},
(Reading: Addison-Wesley)
\bibitem{magorrian}
Magorrian, J. et al. 1998, Astron. J. 115, 2285
\bibitem{mateo}
Mateo, M. 1998, Ann. Rev. Astr. Astrophys. 36, 435
\bibitem{melott}  
  Melott, A.  and  Schramm, D. N. 1985 Astrophys. J. 298, 1 
\bibitem{mohapatra}
Mohapatra, R. N. and Teplitz, V. L. 2000, astro-ph/0001362
\bibitem{moore98}
Moore, B., Governato, F., Quinn, T., Stadel, J., and Lake, G. 1998, Astrophys. J. 499, L5
\bibitem{moore99a}
Moore, B., Ghigna, S., Governato, F., Lake, G., Quinn, T., Stadel, J.,
and Tozzi, P. 1999, Astrophys. J. 524, L19
\bibitem{moore99b}
Moore, B., Quinn, T., Governato, F., Stadel, J., and Lake, G. 1999,
MNRAS, submitted (astro-ph/9903164)
\bibitem{navarrosteinmetz}
Navarro, J. F. and Steinmetz, M. 1999, Astrophys. J., in press (astro-ph/9908114)
\bibitem{nfw}Navarro, J. F., Frenk, C. S. and White, S. D. M. 1996,
MNRAS 462, 563
\bibitem{olszewski}
Olszewski, E. and Aaronson, M. 1987, Astron. J. 94, 657
\bibitem{ostriker}
Ostriker, J. P., astro-ph/9912548, Phys. Rev. Lett., submitted
\bibitem{pagels}
H. Pagels and J.R. Primack 1982,   Phys. Rev. Lett., 48, 223
\bibitem{peacock}
Peacock, J. A. 1999, {\it Cosmological Physics}, (Cambridge: Cambridge University Press)
\bibitem{peebles}
Peebles, P. J. E. 1993, {\it Principles of Physical Cosmology},
(Princeton: Princeton University Press)
\bibitem{pickering}
Pickering, T. E., van Gorkom, J. H., Impey, C. D., and Quillen, A. C. 1999,
Astron. J. 118,765-776
\bibitem{cdm}
Primack, J. R. 1997, in Formation of Structure in the Universe, ed. A. Dekel
and J. P. Ostriker, astro-ph/9707285
\bibitem{primack}
 Primack, J. R.  and   Blumenthal, G.  in
Formation and Evolution of Galaxies and Large Structures in the
Universe, J. Audouze and J. Tran Thanh Van, eds. (Reidel, Dordrecht,
1983) pp. 163-183; reprinted in Beyond the Standard Model,
J.Tran Thanh Van, ed. (Editions Frontieres, 1983) pp. 445-464
\bibitem{ralston}
Ralston, J. P. and Smith, L. L. 1991, Astrophys. J. 367, 54
\bibitem{shi}
Shi, X., and Fuller, G. M. 1999, Phys. Rev. Lett. 82, 2832
\bibitem{sommerlarsen}
Sommer-Larsen, J., and Dolgov, A. 1999, astro-ph/9912166
\bibitem{spergel}
Spergel, D. N. and Steinhardt, P. J. 1999, astro-ph/9909386
\bibitem{swaters}
Swaters, R. 1999, Ph.D. thesis, Groningen
\bibitem{swaters2}
Swaters, R. A., Madore, B. A., and Trewhella, M. 2000, Astrophys. J. in press, astro-ph/0001277
\bibitem{tremaine}
Tremaine, S. and Gunn, J. E. 1979, Phys. Rev. Lett. 42, 407
\bibitem{tyson}
Tyson, J. A., Kochanski, G. P., and Dell'Antonio, I. P. 1998,
Astrophys. J. 498,L107
\bibitem{vandenbosch}
van den Bosch, F. C., Robertson, B. E., Dalcanton, J. J.,
 \& de Blok, W. J. G. 2000, Astron. J. 119, 1579.
\bibitem{wadsley}
Wadsley, J. et al. 1999, in preparation
\bibitem{whiterees}
White, S. D. M., and Rees, M. J. 1978, MNRAS 183, 341
\bibitem{white}
White, M. and Croft, R. A. C. 2000, astro-ph/0001247
\bibitem{williams}
 Williams, L. L. R., Navarro, J. F., Bartelmann, M. 1999,
Astrophys. J. 527,535
\bibitem{zeldovich}
Zeldovich, Ya. B. and Novikov, I.D. 1971,
{\it Relativistic Astrophysics}, vol. 1 (Chicago: University of Chicago press)
\end{thebibliography}
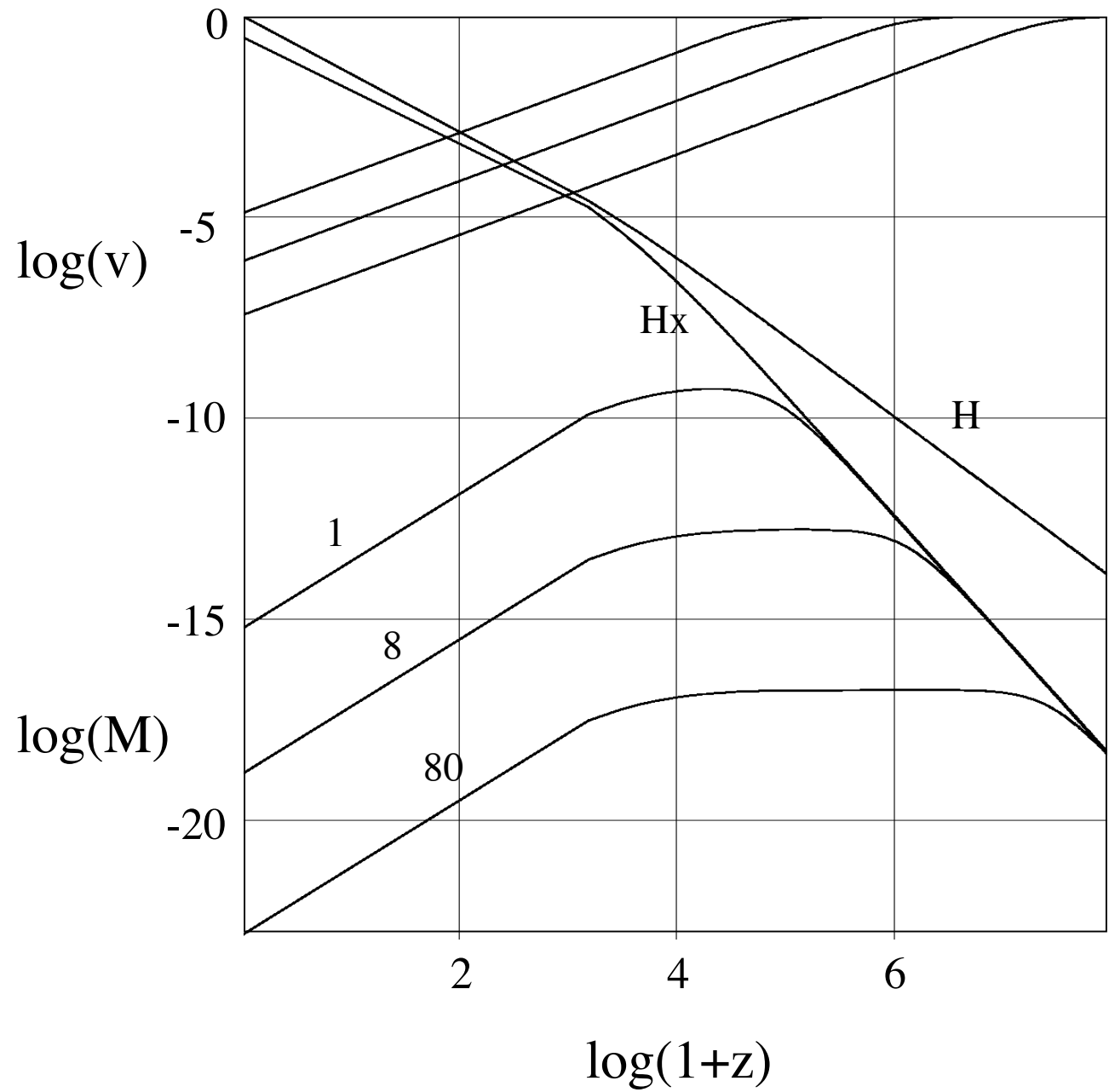
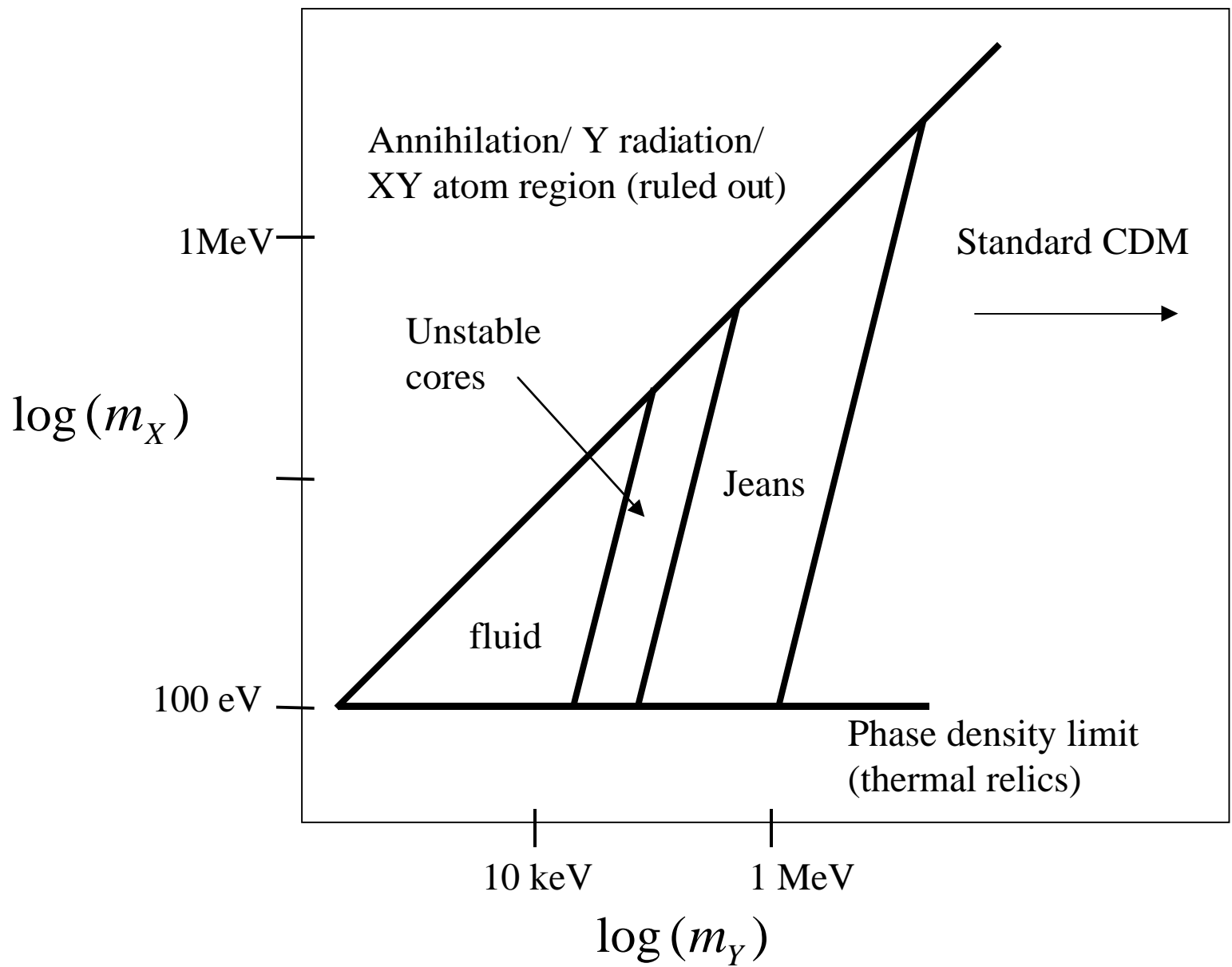
\end{document}